# Can Impurity Effects Help to Identify the Symmetry of the Order Parameter of the Cuprates?

R. Fehrenbacher and M.R. Norman

*Science and Technology Center for Superconductivity, Materials Science Division, Argonne National Laboratory, Argonne, Illinois 60439, USA*

The effects of non-magnetic impurities on the properties of superconductors with order parameters (OP) of different symmetry are discussed. In particular, we contrast the case of a $d_{x^2-y^2}$ with various forms of an anisotropic s-wave (ASW) gap. The biggest qualitative difference occurs if the phase of the s-wave OP does not change sign.

The symmetry of the OP in the high-$T_c$ cuprates has been controversial since the very early days of the field. In this paper, we address the question to what extent the effects of non-magnetic impurities may help to identify this important property. Many recent experiments provide strong evidence for the existence of nodes in the gap. However, line nodes can arise from pairing states with different symmetry. Since it is difficult to map out the **k**-dependence of the gap experimentally, it is important to know whether the symmetry of the OP could be distinguished by the response of the superconductor to the presence of non-magnetic impurities (see also Ref. [1, 2, 3, 4]).

The most serious candidates for the OP seem to be a $d_{x^2-y^2}$ or an ASW gap. Here we consider a two-dimensional electron system with a circular Fermi surface, and a separable pair potential $V_{\mathbf{kk'}} = V\eta(\phi)\eta(\phi')$ ($\phi$ the polar angle in the plane). We consider two possibilities for the angular function $\eta(\phi)$: (i) The dependence $\eta_d(\phi) = \cos 2\phi$, which leads to the characteristic $d_{x^2-y^2}$ gap, $\Delta(\phi) = \Delta \cos 2\phi$, and (ii) $\eta_s(\phi) = |1-4\phi/\pi| - \eta_0$, $0 \leq \phi \leq \pi/2$, periodically continued to the interval $[\pi/2, 2\pi]$, *i. e.,* a sawtooth function with period $\pi/2$. The latter case leads to a particular ASW state with $\Delta(\phi) = \Delta\eta_s(\phi)$. For $\eta_0 = 0$, the gap has a nodal structure identical to the $d_{x^2-y^2}$ case, but no sign change, for $0 < \eta_0 < 1$ it has 8 nodes and a sign change, and for all other values of $\eta_0$ there are no nodes.

We treat the impurity effects in the self-consistent $t$-matrix approximation [5] in Nambu space: The self-energy $\widehat{\Sigma} = \sum_{j=0}^{3} \Sigma_j \widehat{\sigma}_j$ (and all other matrix quantities, marked by a hat) is expanded in terms of the unit and Pauli matrices $\widehat{\sigma}_0 \ldots \widehat{\sigma}_3$ and is given by $\widehat{\Sigma}(i\omega_n) = \Gamma\widehat{T}(i\omega_n)$, where $\widehat{T}$ is the $t$-matrix, $\Gamma = n_i/(\pi N_0)$, $n_i$ the impurity concentration, and $N_0$ the DOS at $\varepsilon_F$ in the normal state. In the present case, the non-vanishing $t$-matrix components are given by

$$T_0 = \frac{G_0}{c^2 - G_0^2 + G_1^2} \qquad T_1 = \frac{-G_1}{c^2 - G_0^2 + G_1^2}. \quad (1)$$

Here we assumed s-wave scattering parametrized in terms of $c = \cot \delta_0$, and $\widehat{G} = (\pi N_0)^{-1} \sum_{\mathbf{k}} \widehat{g}(\mathbf{k})$, where $\widehat{g}(\mathbf{k})$ is the single particle propagator, and $\delta_0$ the phase shift.

The fundamental difference between the two states appears in the presence of non-magnetic impurities, signaled by

$$G_1(i\omega_n) = \frac{1}{\pi N_0} \sum_{\mathbf{k}} \frac{\widetilde{\Delta}_{\mathbf{k}}}{\widetilde{\omega}_n^2 + \widetilde{\Delta}_{\mathbf{k}}^2 + \xi_{\mathbf{k}}^2}. \quad (2)$$

with $\widetilde{\omega}_n = \omega_n + i\Sigma_0$, $\widetilde{\Delta}_{\mathbf{k}} = \Delta_{\mathbf{k}} + \Sigma_1$. Here $\xi_{\mathbf{k}}$ is the quasi-particle energy measured from $\varepsilon_F$. The symmetry of the $d_{x^2-y^2}$ gap makes the sum over **k** vanish, and therefore $G_1 = \Sigma_1 = 0$, *i. e.*, $\widetilde{\Delta}_{\mathbf{k}} = \Delta_{\mathbf{k}}$. In the ASW state, $G_1$ vanishes only at $\eta_0 = 0.5$ (in which case there is no qualitative difference to the $d_{x^2-y^2}$ state), but in general, $G_1$ and $\Sigma_1$ are finite. A finite $G_1$ leads to an *isotropic* renormalization of the off-diagonal self-energy, $\widetilde{\Delta}_{\mathbf{k}} = \Delta_{\mathbf{k}} + \Sigma_1$. Below, we shall discuss how the gap renormalization affects the density of states (DOS) $N(\omega)$ at low energies.

In Fig. 1, we compare the DOS (which is defined by $N(\omega)/N_0 = -\Im m G_0(i\omega_n)|_{i\omega_n=\omega+i\delta}$) for the following gap functions at various impurity concentrations ($\Gamma$ is measured in units of $\Delta_{00}$, the pure gap at $T = 0$) in the unitary limit ($c = 0$): (a) The $d_{x^2-y^2}$ gap, and (b) the ASW gap with $\eta_0 = 0, 0.1, 0.4$. The pure DOS is linear at small frequencies for all states. In the presence of impurities, the $d_{x^2-y^2}$ state shows the characteristic zero frequency resonance leading to a large DOS at $\varepsilon_F$. The ASW states show different features: For $\eta_0 = 0$ a gap opens for any finite impurity concentration as shown previously [2]. This is due to the fact that the pure gap does not change sign, hence the mixing of Cooper pairs with different **k** by the s-wave scattering immediately results in a finite gap at all **k**. For $\eta_0 = 0.1$, a finite DOS at $\varepsilon_F$ appears for small $\Gamma$, but eventually, at larger $\Gamma$, the mixing of the pairs averages out the regions of negative $\Delta$, and a gap appears, which in the limit $\Gamma \to \infty$ becomes equal to the Fermi surface average of the pure gap. The latter is true for all ASW states. For small $\eta_0 \ll 0.5$, the impurity concentration necessary to produce a gap is approximately given by $\Gamma_{\text{crit}} \approx \eta_0(c^2+1)$. At this value, the effective normal state inverse scattering rate $\Gamma/(c^2+1)$ is large enough to cancel the sign change of the pure OP. For $\eta_0 = 0.4$, the situation is even more complex: now the zero frequency resonance is not completely cut off by the off-diagonal self-energy (as a result of the small value of the Fermi surface average of the pure OP), and leads to a large DOS at $\varepsilon_F$, already for small $\Gamma$ analogous to the $d_{x^2-y^2}$ state. Finally, at very large $\Gamma$, a small gap forms.

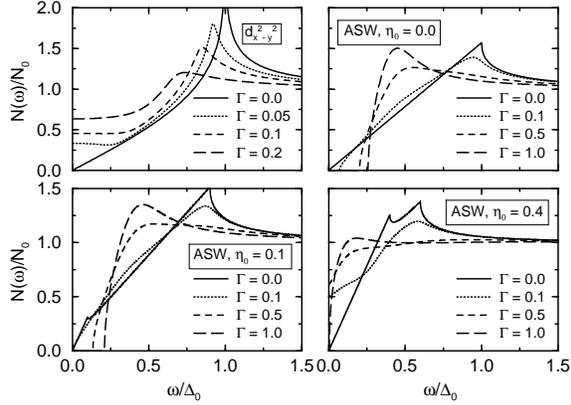

**FIG.1**. A comparison of the DOS at $T = 0$ for the $d_{x^2-y^2}$ OP and different ASW OPs in the unitary limit ($c = 0$) at different values of the impurity concentration ($\Delta_0$ is the gap at $T = 0$ in the presence of impurities).

In Fig. 2, we compare the low frequency DOS in the Born and unitary limit for ASW states with different values of $\eta_0$ close to the 'critical' value $\eta_0 = 0.5$ at small impurity concentration (we chose $\Gamma_{c=1} = 2\Gamma_{c=0}$, so that the normal state scattering rates are the same).

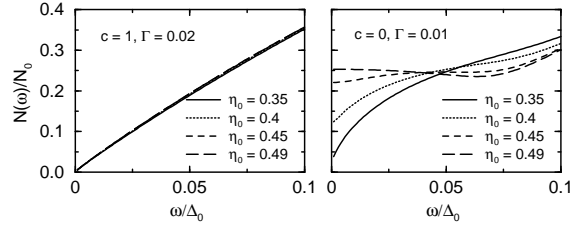

**FIG.2**. A comparison of the low energy DOS at $T = 0$ in the Born ($c = 1$) and unitary limit for ASW OPs with different values of $\eta_0$ at small impurity concentration ($\Delta_0$ is the gap at $T = 0$ in the presence of impurities).

In the unitary limit, a large residual DOS at $\varepsilon_F$ similar to the $d_{x^2-y^2}$ state is found down to values $\eta_0 = 0.35$. This comparison clearly shows that the qualitative difference between the two limits persists into a wide range of parameters $\eta_0$ away from $\eta_0 = 0.5$, at which the gap average vanishes.

In conclusion, our results show that based on quantities which probe the DOS at low energies, it is very difficult to distinguish between a $d_{x^2-y^2}$ and an ASW state. If the OP of the ASW state does not change sign, a finite gap always exists in the presence of impurities, and therefore the distinction is clear. However, as we have shown, if the ASW gap is allowed to change sign, there is a smooth transition from a behaviour completely analogous to the $d_{x^2-y^2}$ state at $\eta_0 = 0.5$ to the case of a gap formation at $\eta_0 = 0.0$. These observations suggest that impurity effects can be used to identify the symmetry of the OP *only* if additional information on the **k**-dependence of its modulus is known, *e. g.,* from angle-resolved photoemission. This emphasizes the importance of phase-sensitive experimental probes to resolve this issue.

We acknowledge the financial support of the National Science Foundation through grant NSF-DMR-91-20000 (R.F.), and of the U.S. Department of Energy, Office of Basic Energy Sciences, under Contract No. W-31-109-ENG-38 (M.R.N.).